\documentclass[aps,prb,twocolumn,floatfix]{revtex4}

\usepackage{epsfig}
\usepackage{graphicx}
\usepackage{amsfonts}
\usepackage{epstopdf}
\begin{document}

\title{Self-assembly, Structure and Electronic Properties of a Quasiperiodic Lead Monolayer}

\author{J. Ledieu\footnote{Corresponding author:
e-mail: Ledieu@lsg2m.org}}
\affiliation{ LSG2M, CNRS UMR 7584, Ecole des Mines, Parc de Saurupt, 54042 Nancy Cedex, France\\}
\author{L. Leung\footnote{Present address:
Department of Chemistry, University of Toronto, Toronto, Ontario M5S 3H6, Canada}, L.H. Wearing, R. McGrath}
\affiliation{\textit{Surface Science Research Centre and Department of
Physics, \\ The University of Liverpool, Liverpool L69 3BX, UK\\}}
\author{T.A. Lograsso, D. Wu}
\affiliation{\textit{Ames Laboratory, Iowa State University, Ames, IA 50011, USA\\}}
\author{V. Fourn\'{e}e}
\affiliation{\textit{ LSG2M, CNRS UMR 7584, Ecole des Mines, Parc de Saurupt, 54042 Nancy Cedex, France\\}}


\begin{abstract}
A quasiperiodic Pb monolayer has been formed on the five-fold surface of the Al-Pd-Mn quasicrystal. Growth of the monolayer proceeds via self-assembly of an interconnected network of pentagonal Pb stars, which are shown to be $\tau$-inflated compared to similar structural elements of the quasiperiodic substrate. Measurements of the electronic structure of the system using scanning tunnelling spectroscopy and ultra-violet photoemission spectroscopy reveal that the Pb monolayer displays a pseudo-gap at the Fermi level which is directly related to its quasiperiodic structure.

\end{abstract}

\maketitle

Quasicrystals are complex intermetallic alloys with long-range aperiodic order and non-crystallographic rotational symmetry \cite{Shechtman84}. The physical properties arising from the quasiperiodic arrangement of metal atoms in quasicrystals significantly depart from those of periodic alloys \cite{Trebin03}. The most surprising feature is perhaps the fact that although quasicrystals are alloys of metallic elements, they behave as poorly metallic systems. A decrease of the spectral intensity at the Fermi level corresponding to the presence of a pseudo-gap \cite{Stadnik01} in the density of states (DOS) is found in bulk electronic measurements of quasicrystals. The pseudogap arises because of the Fermi surface - Brillouin zone interaction and it has been shown that the transport properties are dependent on the depth of the gap \cite{Mizutani98}. Consequently, there is considerable interest in clarifying the relative importance of quasiperiodic order and the complex chemistry of quasicrystals in determining the width and depth of the pseudogap.

The issue would be considerably simplified if there were single element quasicrystals. As such materials do not exist in bulk form,  there have been several  recent attempts to grow single element quasiperiodic systems using quasicrystalline surfaces as templates. A number of elements with differing chemistry have been utilised in an attempt to achieve this goal. On both Al-Pd-Mn \cite{Ledieu06} and Al-Ni-Co \cite{Leung06}, Si atoms have been found to occupy unique sites at sub-monolayer coverages, leading to ordered adsorption; however at coverages $>0.33$ monolayers (ML), the structures became increasingly disordered due to occupation of a range of other adsorption sites. Aluminum atoms were found to order into pentagonal ``starfish" on the five-fold surface of Al-Cu-Fe at low submonolayer coverages, though this initial ordering did not lead to the formation of ordered monolayers \cite{Cai01}. Franke \textit{et al.} first demonstrated quasiperiodicity in Bi and Sb monolayers on Al-Pd-Mn and Al-Ni-Co quasicrystal surfaces \cite{Franke03}. These measurements were done using diffraction techniques, which do not yield information on how such layers assemble. Moreover no quantitative information on either the structure of the monolayers or their electronic properties was obtained.

In this paper we present new findings on the self-assembly, structure and electronic properties of a quasiperiodic monolayer. We have used the  five-fold Al-Pd-Mn surface as a template for Pb adsorption at a range of fluxes and substrate temperatures. The initial growth and stability of the monolayer was studied by scanning tunneling microscopy (STM). The structure of the Pb film is found to be $\tau$-inflated compared to the quasiperiodic substrate, where $\tau$, the golden mean, is an irrational number (=1.618...) associated with pentagonal geometry. The electronic structure of the system was measured by ultra-violet photoemission (UPS) and scanning tunneling spectroscopy (STS), and it is demonstrated unambiguously that the aperiodic structure influences the electronic density of states such that the Pb monolayer exhibits a pseudogap at the Fermi level.

These experiments were performed under ultra high vacuum using low energy electron diffraction (LEED), Auger electron spectroscopy (AES), x-ray and ultra violet photoemission spectroscopy (XPS and UPS respectively), scanning tunneling spectroscopy (STS) and variable temperature scanning tunneling microscopy (VT-STM). The  Al$_{70}$Pd$_{21}$Mn$_{9}$ sample was cut perpendicular to its five-fold symmetry axis. Lead deposition was carried out using an electron beam cell with a flux of  \mbox{2.5 x 10$^{-3}$ML/s}. The Pb source was calibrated by means of XPS and STM using an Al(111) crystal.

\begin{figure}
 \begin{center}
\includegraphics[width=0.46\textwidth]{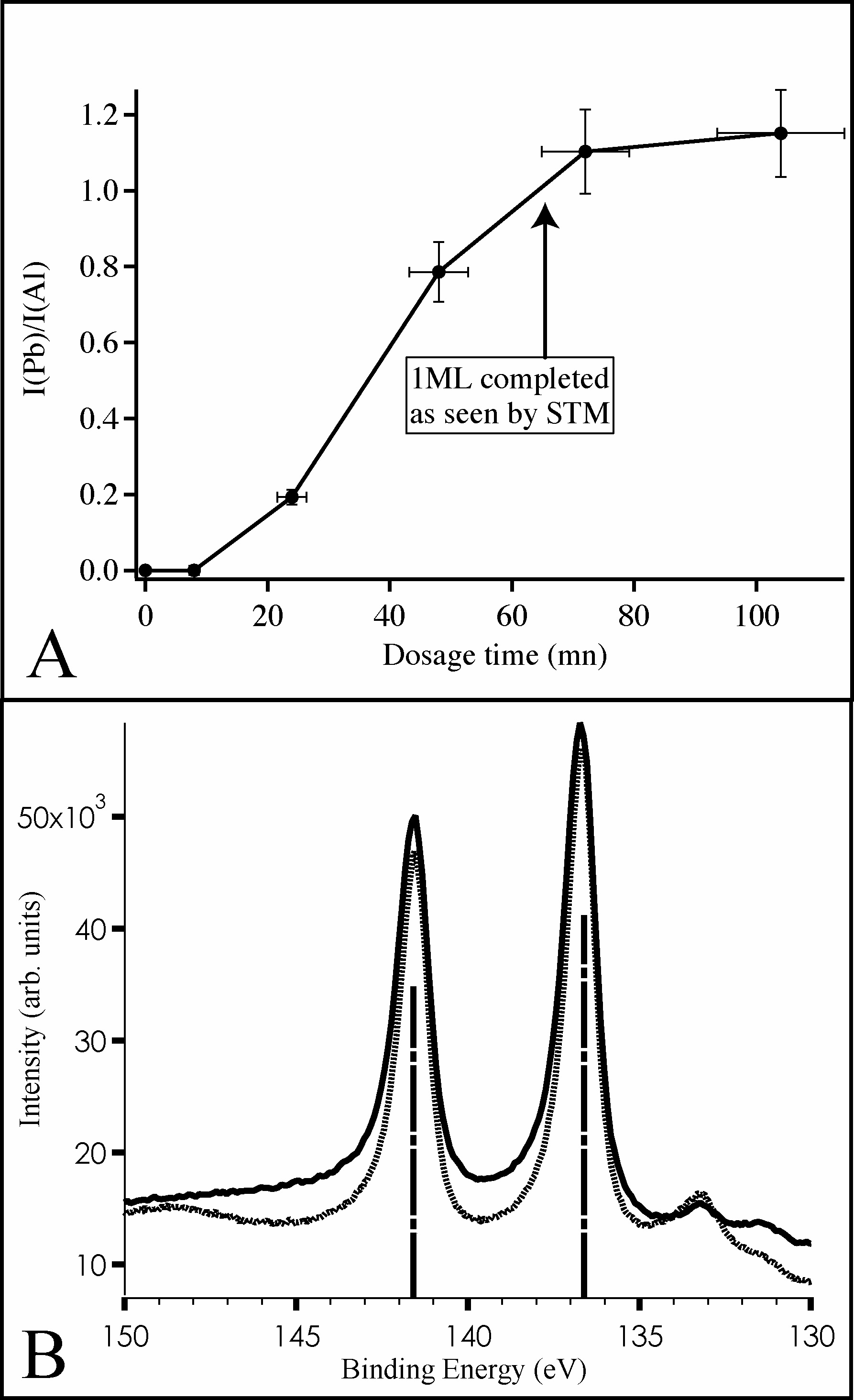}
 \caption{a) Ratio of the intensities of the Pb$_{NOO}$ (96 eV) and Al$_{LVV}$ (68 eV) Auger peaks as a function of the coverage in dosage time. b) XPS spectra of the Pb 4\textit{f} core levels for 1ML of Pb adsorbed on the five-fold Al-Pd-Mn quasicrystal surface (full curve) at 653 K and on Al(111) (dashed line). The markers at 136.6 eV and 141.5 eV indicate the position for elemental Pb.}
  \label{fig1}
 \end{center}
 \end{figure}

We first outline the adsorption characteristics of the adsorption of Pb as monitored by AES and XPS. The evolution of the intensity ratio of the Pb$_{NOO}$ (96 eV) and Al$_{LVV}$ (68 eV) Auger peaks as a function of  dosage time is plotted on Fig.\ref{fig1}(a). The rate of adsorption of Pb decreases with increasing Pb coverage; the curve initially rises exponentially and then reaches saturation at monolayer coverage. No  further growth was observed under the range of fluxes (\mbox{2.5 x 10$^{-3}$ML/s} - \mbox{2.5 x 10$^{-2}$ML/s}) and substrate temperatures (57 K - 653 K) used in these measurements. The film was observed to desorb totally at a temperature of 670 K which is higher than the bulk melting point of Pb. The Al $2p$ and Pb $4f$ core levels were also recorded upon adsorption. The shape of the Al $2p$ core level measured is identical to that of the clean quasicrystal surface for the monolayer as deposited at room temperature or annealed to 653 K. The Pb $4f$ core level recorded from the quasiperiodic Pb monolayer has an identical shape and intensity to that measured from 1ML of Pb grown on Al(111) (see Fig.\ref{fig1}(b)). The lack of new components and chemical shifts within the core level peaks is consistent with the immiscibility of Pb and Al \cite{Johnson01,Matolin06}.

\begin{figure}
 \begin{center}
\includegraphics[width=0.48\textwidth]{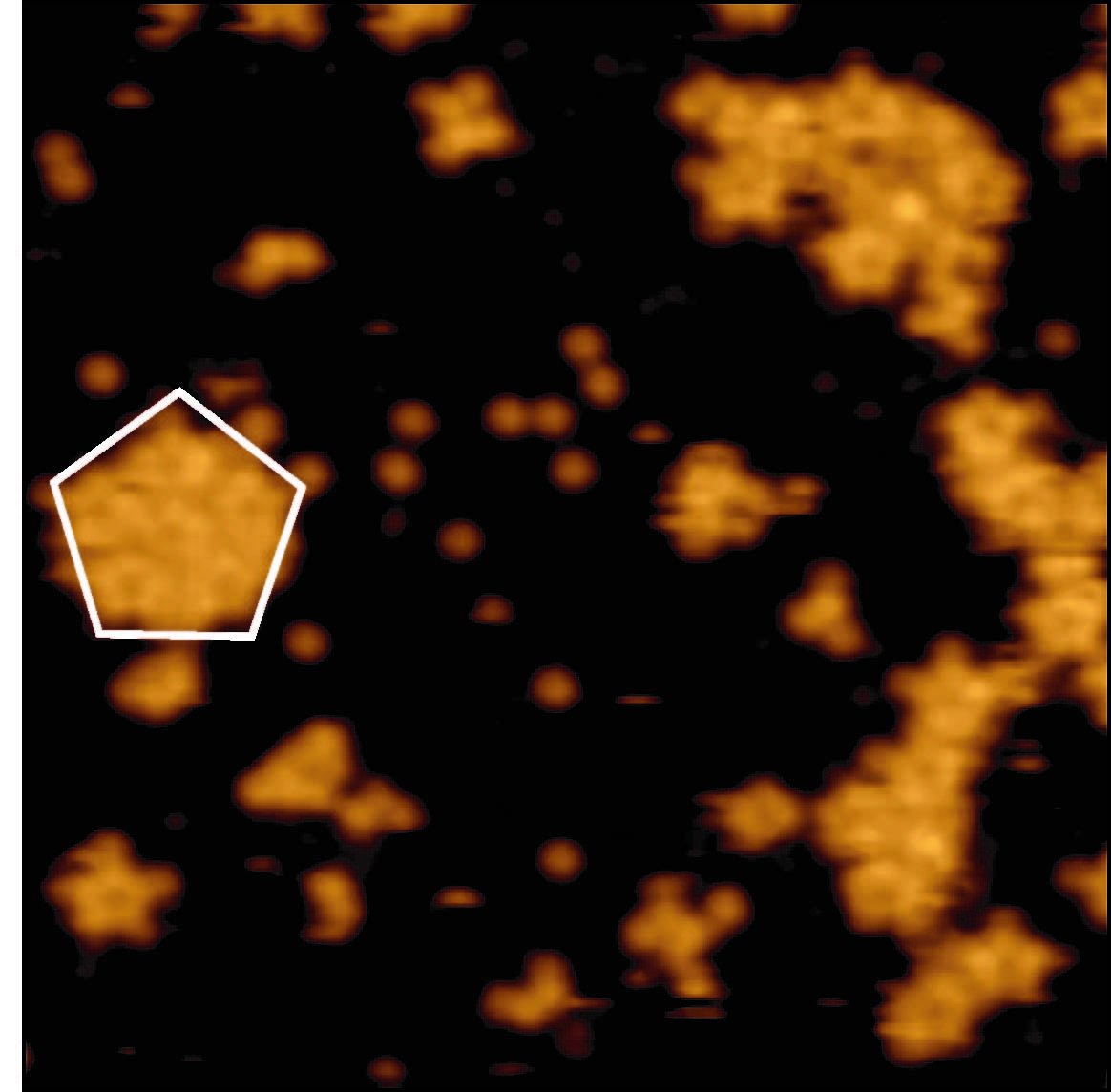}
 \caption{Color online: 180 \AA{} x 180 \AA{} STM image for 0.2 ML of Pb adsorbed on the five-fold surface of Al-Pd-Mn. A large pentagon formed by five smaller pentagonal islands is outlined.}
  \label{fig2}
 \end{center}
 \end{figure}

We now describe observations of the self-assembly of the Pb monolayer at the atomic level. Fig.\ref{fig2} shows the Al-Pd-Mn surface after the adsorption of 0.2 monolayer (ML) of Pb.   At this coverage, individual Pb atoms, incomplete pentagonal stars, complete pentagonal stars and networks of pentagonal stars are all visible on a single terrace.  The formation of these networks indicates that the Pb atoms are mobile enough to diffuse on the surface \cite{Fournee05,Ledieu04}. We have observed this network formation in the temperature range 57-653 K. With increasing coverage, the density of stars increases with additional atoms filling the interstices within the network. The edge length of the pentagon formed by joining the tips of these pentagonal stars  is $4.9\pm0.3$\AA{} and they are mono-atomic in height. The smallest pentagonal structural elements present on the clean Al-Pd-Mn surface have an edge length of 3.0\AA{} \cite{ledieu02}, i.e $\tau$ smaller than the smallest pentagon within the Pb structure.  This $\tau$-scaling of the basic structural elements is reflected in a $\tau$-scaling of the complete monolayer structure (described below).

The pentagonal stars all have the same orientation on this terrace and on adjoining terraces, indicating that they nucleate at unique sites.
Pentagonal islands of a similar size and orientation were  reported  by Cai \textit{et al.} when dosing Al on the isostructural Al-Cu-Fe surface \cite{Cai01}.   Both Al and Pb pentagonal islands are obtained regardless of the deposition rate, a fact characteristic of a heterogeneous nucleation process at specific trap sites \cite{Cai01,Fournee03,Evans06}. However, Al starfish islands do not develop into a quasiperiodic layer upon further deposition, which is indicative of different kinetics in that system.

\begin{figure}
 \begin{center}
\includegraphics[width=0.48\textwidth]{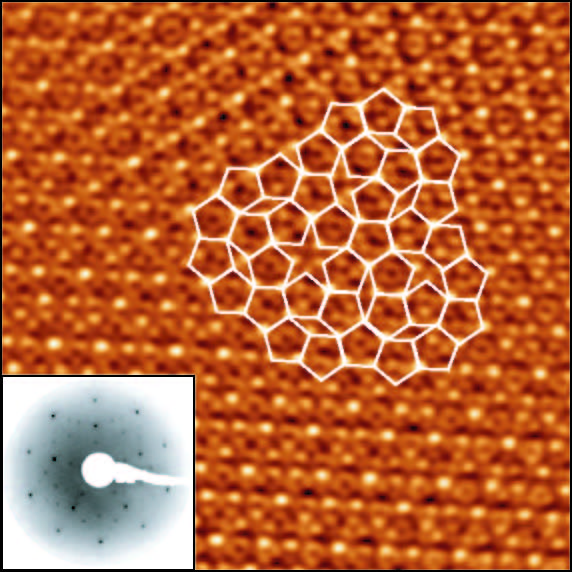}
 \caption{Color online: 250 \AA{} x 250 \AA{} STM image of 1.0 ML of Pb adsorbed on the five-fold surface of Al-Pd-Mn. Inset: LEED pattern recorded at 80 eV at the same coverage.}
  \label{fig3}
 \end{center}
 \end{figure}

The structural characteristics of the completed Pb monolayer are now described. The complete monolayer exhibits a well-ordered quasiperiodic structure as evidenced by the five-fold LEED pattern (inset in Fig.\ref{fig3}) and from the analysis of atomically resolved STM images.  The  monolayer density is 0.09 atom/\AA$^{2}$, deduced from the intensity of the Pb 4f core-level compared to that measured of a Pb monolayer deposited on an Al(111) surface (see Fig.\ref{fig1}(b)). It exists two equivalent ways to improve the overall structural quality of the Pb monolayer; either by depositing at room temperature then annealing at 653 K or by depositing Pb while maintaining the substrate at 653 K . Fig.\ref{fig3} shows an STM image acquired following the latest procedure which is used for the rest of the paper. The roughness of the film Z$_{rms}$ is 0.2\AA{}, comparable to that of the quasicrystal substrate \cite{ledieu02}. A Fast Fourier Transform (FFT) calculated on a larger STM image (not shown here) exhibits ten-fold symmetry, and is comparable to FFTs calculated from STM images obtained from the clean five-fold Al-Pd-Mn substrate except for a $\tau^{-1}$ scaling of distances in reciprocal space. A patch of a Penrose (P1) tiling superimposed on the Pb monolayer emphasizes the quasiperiodic nature of the deposited monolayer (Fig.\ref{fig3}). The tiling is composed of four tiles: the regular pentagon, the rhombus, the crown and the  pentagonal star. A similar tiling was derived from the clean five-fold Al-Pd-Mn surface \cite{ledieu02}, but the tiles used on Fig.\ref{fig3} are $\tau$ inflated compared to those found for the clean surface. This $\tau$-scaling in the basic structure is also evident from a simple inspection of LEED patterns.

\begin{figure}
 \begin{center}
 \includegraphics[width=0.48\textwidth]{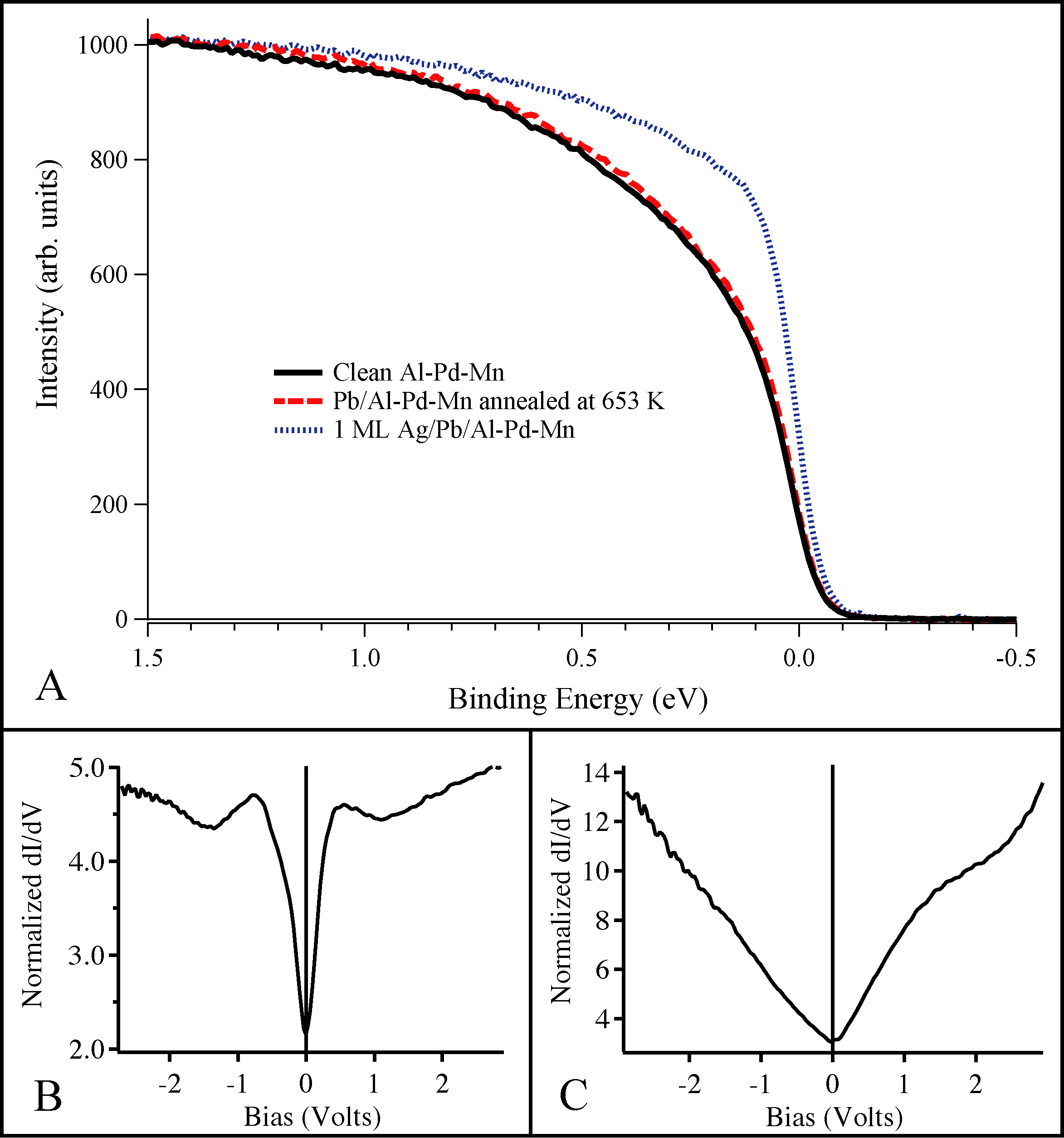}
  \caption{Color online: a) UPS measurements at the Fermi level for several systems: clean surface (black, full line), a monolayer of Pb annealed on the quasicrystal (red, large dash), and with an additional Ag monolayer deposited (blue, small dash). b) and c) STS measurements on the clean and Pb dosed five-fold Al-Pd-Mn surface.}
   \label{fig4}
\end{center}
 \end{figure}

The electronic structure of the monolayer was probed using UPS and STS. Fig.\ref{fig4}(a) presents UPS measurements close to the Fermi level for the clean Al-Pd-Mn quasicrystal surface, an annealed Pb monolayer on the quasicrystal surface, and 1ML of Ag adsorbed on an annealed Pb monolayer on a quasicrystal surface. Using a previously established procedure \cite{Stadnik01}, the depth $C$ of the pseudo-gap was measured for the three systems. In this procedure the depth parameter  $C=0 \%$ is for a metallic sample and $C=100 \%$ for an insulator. Although a value of $C= 35 \% $ is found for both quasiperiodic structures, the depth is reduced to 16$ \% $ after deposition of a disordered Ag monolayer on the Pb thin film. This demonstrates that a single metallic layer is sufficient to greatly enhance the spectral intensity in the vicinity of the Fermi level.

In addition to UPS, STS measurements were collected on the annealed Pb monolayer and the clean Al-Pd-Mn surface (see Fig.\ref{fig4}(b,c)). Normalised \emph{dI/dV} spectra represent an average over 10$^{4}$ I(V) curves before differentiation. These results are tip independent and  I(V) curves were collected in a grid-like manner on different local atomic configurations. The pseudo-gap appears larger on the Pb structure (Fig.\ref{fig4}(c)) than on the clean surface (Fig.\ref{fig4}(b)). This effect is consistent with the UPS analysis, as the UPS signal includes a contribution from the substrate to the spectral intensity at E$_{F}$;  STS measurements probe only the DOS of the top-most surface layer hence reducing the Al-Pd-Mn electronic contribution to the signal detected. Furthermore, in a nearly free electron like system, the width of the gap scales with the lattice potential  $\vert\textit{V$_{K}$}\vert$ which is the product of the form factor \textit{w$_{K}$}  and the geometrical structure factor \textit{S$_{K}$} of the crystal \cite{Duboisbooks,Fournee02}.  The larger gap measured by STS on the Pb film compared to the clean Al-Pd-Mn surface is expected, as the form factor \textit{w$_{K}$}  is directly related to the atomic number of the element probed. With an identical crystallographic structure for both the substrate and the deposited monolayer, the heaviest element should open a larger gap. We infer that the quasiperiodic structure is responsible for the formation of this pseudo-gap.

In conclusion, we have successfully used the Al-Pd-Mn surface as a template to grow a monoelement quasiperiodic monolayer. The monolayer self-assembles via a network of pentagonal islands. The structure of the monolayer is $\tau$-inflated compared to the surface of the substrate. The reduced chemical complexity allow us to correlate the quasiperiodic structure with the formation of a pseudo-gap in the DOS at E$_{F}$. The system represents an interesting testing ground for further experimental investigations on the physical properties associated with aperiodic order.

We acknowledge the European Network of Excellence on Complex Metallic Alloys (CMA) contract NMP3-CT-2005-500145, EPSRC (grant number GR/S19080/01) and the U.S.  Department of Energy and the Basic Energy Sciences for financial support.  We also thank G. Darling for fruitful discussions.

\end{document}